\documentclass[a4paper, fontsize=10pt]{article}
\usepackage[utf8]{inputenc}
\usepackage[top=25mm, bottom=25mm, left=25mm, right=25mm]{geometry}
\usepackage{graphicx}
\usepackage{epstopdf}
\usepackage{amssymb}
\usepackage{amsmath}
\usepackage[hidelinks]{hyperref}
\usepackage{float}
\usepackage[super, numbers, comma, sort&compress]{natbib}
\usepackage{microtype}
\usepackage{marginnote}
\usepackage[usenames, dvipsnames]{xcolor}
\usepackage{colortbl}
\usepackage{multirow}
\usepackage{lipsum}
\usepackage{fancyhdr}
\usepackage{longtable}
\usepackage{soul}
\usepackage{eurosym}
\usepackage[shortlabels]{enumitem}
\usepackage{tcolorbox}
\usepackage[hang, flushmargin]{footmisc}
\usepackage{titlesec}
\usepackage[version=4,arrows=pgf-filled,textfontname=sffamily]{mhchem}
\usepackage[font=small]{caption}
\usepackage{siunitx}
\usepackage[bbgreekl]{mathbbol}
\usepackage{bm}
\usepackage{lipsum}
\usepackage{authblk}
\usepackage{abstract}
\usepackage{placeins}

\DeclareSymbolFontAlphabet{\mathbbm}{bbold}
\DeclareSymbolFontAlphabet{\mathbb}{AMSb}

\titleformat{\section}{\normalfont\Large\bfseries\MakeUppercase}{}{0em}{}
\titleformat{\subsection}{\normalfont\large\bfseries}{}{0em}{}
\renewcommand{\refname}{References}
\titleformat{\refname}{\normalfont\Large\bfseries\MakeUppercase}{}{0em}{}

\begin{document}


\title{\vspace{-20pt} Tunable self-emulsification via viscoelastic control of Marangoni-driven interfacial instabilities}
\author[1,2]{Christoph Haessig}
\author[3]{Mehdi Habibi}
\author[1]{Uddalok Sen\footnote{uddalok.sen@wur.nl, ORCID: \href{https://orcid.org/0000-0001-6355-7605}{0000-0001-6355-7605}}}
\affil[1]{Physical Chemistry and Soft Matter group, Wageningen University and Research, 6708 WE Wageningen, The Netherlands}
\affil[2]{Department of Chemical Engineering, KU Leuven, 3001 Leuven, Belgium}
\affil[3]{Laboratory of Physics and Physical Chemistry of Foods, Wageningen University and Research, 6708 WG Wageningen, The Netherlands}
\date{}
\maketitle


\begin{abstract}
	\vspace{-40pt}
	Interfacial instabilities in multicomponent fluidic systems are widespread in nature and in industrial processes, yet controlling their dynamics remains a challenge. Here, we present a strategy to actively tune Marangoni-driven self-emulsification at liquid-liquid interfaces by harnessing fluid viscoelasticity. When a water-alcohol droplet spreads on an oil bath, a radial surface tension gradient induced by selective alcohol evaporation drives an interfacial instability, leading to the spontaneous formation of a dense two-dimensional array of ``daughter" droplets. We demonstrate that introducing trace amounts of high-molecular-weight polymers, which introduces viscoelasticity, provides a robust means of controlling this process. Increasing viscoelasticity systematically suppresses the instability, resulting in a delayed onset of fragmentation and longer spreading fingers. By combining high-resolution experimental visualization and theoretical analysis, we uncover a quantitative relationship between the polymer concentration and the finger length prior to breakup. These findings establish a predictive framework for designing viscoelastic interfacial materials with programmable dynamics and offer new opportunities for surface-tension-mediated patterning, emulsification, and fluidic control in soft material systems. 
\end{abstract}



\section{Introduction}

Controlling interfacial dynamics is central to a wide range of functional material systems, including coatings \cite{weinstein-2004-arfm, snoeijer-2013-arfm}, inkjet printing \cite{lohse-2022-arfm}, and surface patterning technologies \cite{rauscher-2008-arms}. In this context, the deposition of a droplet onto a liquid substrate presents a particularly rich platform for studying complex interfacial behavior, where morphological alterations---such as the formation of a liquid lens---are driven by the interplay between surface tension, fluid composition, and evaporation \cite{book-degennes, nepomnyashchy-2021-curropincolloidinterfacesci}. In multicomponent droplets, such as water-alcohol mixtures, preferential evaporation of the more volatile component induces radial surface tension gradients, which in turn generate strong Marangoni flows \cite{lohse-2020-natrevphys}. These flows give rise to a myriad of interfacial instabilities, including film spreading \cite{berg-2009-pof}, dewetting \cite{lambooy-1996-prl, oron-2004-prl}, fingering \cite{roche-2014-prl, leroux-2016-pre, ma-2023-jfm}, pattern formation \cite{hack-2024-jcis, chan-2024-pnasnexus}, and even catastrophic topological changes such as ligament break-up and interfacial bursting \cite{yamamoto-2015-natcommun, keiser-2017-prl, wodlei-2018-natcommun, hasegawa-2021-pof, seyfert-2022-prf}. \\

Beyond their undoubted scientific richness as well as visual beauty \cite{lohse-2020-natrevphys}, Marangoni-driven phenomena at liquid-liquid interfaces can also be harnessed as a novel methodology to shape liquid interfaces towards practical applications. One particularly striking example is the so-called ``Marangoni bursting" phenomenon \cite{yamamoto-2015-natcommun, keiser-2017-prl, wodlei-2018-natcommun, hasegawa-2021-pof, seyfert-2022-prf, jaberi-2023-prf}, where a volatile aqueous droplet---typically containing a short-chain alcohol (e.g. 2-propanol, henceforth referred to as ``IPA")---is gently deposited on an immiscible non-volatile oil (e.g. sunflower oil) substrate. The droplet spontaneously spreads on the surface of the oil bath (since the spreading parameter, $S$, is positive \cite{book-degennes}), and forms a liquid lens (see prior studies \cite{keiser-2017-prl, seyfert-2022-prf} for a detailed description of the Marangoni bursting phenomenon). The preferential depletion of alcohol close to the edge of the liquid lens due to evaporation locally raises the interfacial tension, thus driving an outward solutal Marangoni flow \cite{lohse-2020-natrevphys} from the center of the liquid lens towards its edge. However, this increase of interfacial tension is also associated with a concomitant decrease of the spreading parameter, $S$, which eventually becomes negative. This negative spreading parameter now results in a dewetting of the edge of the drop, and competes with the outward Marangoni flow, leading to the formation of a thicker rim at the periphery of the drop. The thick rim destabilizes via a Rayleigh-Plateau-like or contact line instability mechanism \cite{bonn-2009-rmp, keiser-2017-prl}, leading to the spontaneous generation of hundreds of daughter droplets suspended as a two-dimensional array on the oil phase. This self-emulsification process represents a novel and powerful surface-tension-driven strategy for emulsification without external forcing, thus offering considerable promise for patterning, encapsulation, and droplet microfabrication \cite{book-bibette}.  \\

Despite its potential, controlling the dynamics of Marangoni bursting remains a significant challenge. Recent work aimed at using this phenomenon to fabricate functional microstructures (e.g. optical fibres of organic chromophores \cite{slemp-2024-jphyschemc}) emphasizes the need for process-level tunability---particularly in terms of emulsification-onset timing, droplet size, and fragmentation behavior. These dynamics are strongly governed by the morphology of the spreading droplet, particularly the formation and stretching of peripheral fingers that eventually pinch-off into daughter droplets \cite{villermaux-2020-jfm}. \\ 

Here, we propose a materials-based strategy for controlling this spontaneous interfacial instability: the introduction of minute amounts of polymers to impart viscoelasticity to the spreading droplet. The stabilizing effects of polymers on thinning liquid threads are well established in the context of jet breakup, droplet formation and microfluidic stability \cite{middleman-1965-chemengsci, goldin-1969-jfm, sen-2021-jfm}. Viscoelasticity, arising from the relaxation of stretched polymer chains that are dissolved within the liquid, resists deformation and retards thinning---enabling greater control over interface-driven breakup phenomena \cite{eggers-2008-rpp, sen-2021-jfm}. Polymer additives, in minute quantities, have already proven to be an effective control strategy in fields ranging from inkjet printing \cite{sen-2021-jfm} to pesticide treatments \cite{gaillard-2022-jnnfm} to airborne disease transmission \cite{li-2025-arxiv}. While prior studies \cite{ma-2020-pof, motaghian-2022-jcis, ma-2023-jfm} have reported the influence of viscoelasticity on fingering morphologies in Marangoni-driven spreading systems, its use as a tunable parameter to modulate self-emulsification dynamics in Marangoni bursting has not yet been demonstrated. \\

In this work, we experimentally investigate how polymer-induced viscoelasticity can be leveraged to control the complex fragmentation dynamics of Marangoni bursting. Using a model system---water-IPA droplets containing small amounts of dissolved polyethyelene oxide (PEO) of controlled molecular weight and concentration---we show that viscoelasticity delays the onset of bursting, increases the finger length, and alters the instability wavelength. A scaling law is proposed that quantitatively relates the finger stretching dynamics to the elastic properties of the fluid. Our findings offer a robust and generalizable framework for the design of responsive interfacial materials \cite{pujar-2019-langmuir, atis-2019-prx, zhang-2020-acsapplnanomater, choi-2020-jmaterchemc, vdweijden-2020-natcommun, peddireddy-2021-pnas, pascual-2021-prf, yoo-2022-advmater, winkens-2022-langmuir, chen-2023-advfunctmater, winkens-2023-small, devisser-2024-small, korevaar-2025-angewchem} with programmable self-emulsification behavior \cite{tang-2017-advmaterinterfaces, li-2019-acsami, chao-2020-sm, thakur-2021-acsami, mcbride-2024-acsami}, thus bridging fundamental fluid dynamics with functional material design. 

\section{Results}

\subsection{Spontaneous self-emulsification of viscoelastic droplets}

\begin{figure}
    \centering
    \includegraphics[width=\textwidth]{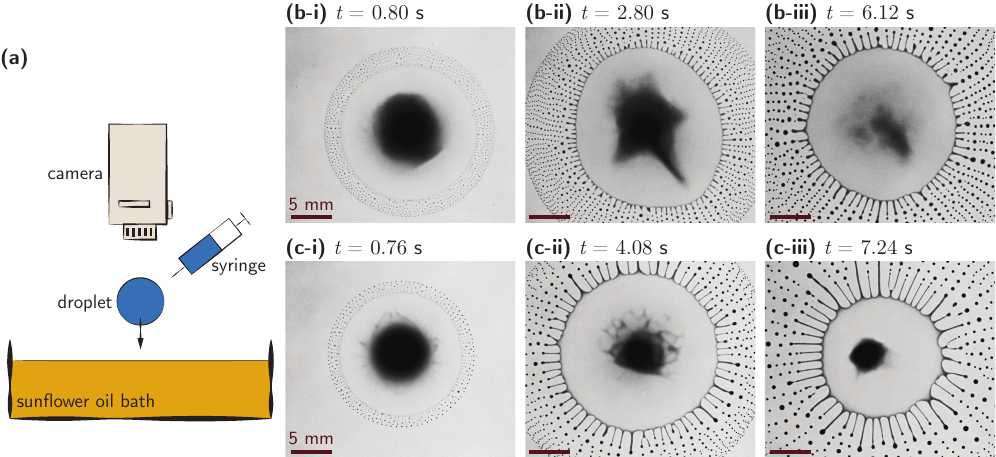}
    \caption{\textbf{Experimental phenomena.} \textbf{(a)} Schematic of experimental setup. Typical time-lapsed snapshots of spontaneous self-emulsification of a water-IPA droplet containing \textbf{(b-i) -- (b-iii)} 0.10\% PEO1M and \textbf{(c-i) -- (c-iii)} 0.50\% PEO1M; the scale bars denote 5 mm. See movies SM1 and SM2 in the Supplementary Information for the corresponding movies.}
    \label{fig:fig-1}
\end{figure}

To investigate the role of viscoelasticity in Marangoni-driven self-emulsification, we deposit a 7.5 \SI{}{\micro \text{L}} droplet of an aqueous 2-propanol (IPA) solution containing dissolved polyethylene oxide (PEO) at mass concentrations, $C_{\mathrm{m}}$, using either PEO1M or PEO4M as the polymer additive. The droplet is gently placed on a quiescent sunflower oil bath and imaged from above using high-speed videography (as shown in figure \ref{fig:fig-1}a; see the Methods section for experimental details). The moment of first contact between the droplet and the oil bath is set as time $t = 0$. \\

Immediately upon contact, the droplet undergoes rapid spreading followed by spontaneous self-emulsification, where it disintegrates into thousands of daughter droplets within a few seconds. A closer inspection of this self-emulsification behavior, as depicted in figures \ref{fig:fig-1}b-i -- \ref{fig:fig-1}b-iii and \ref{fig:fig-1}c-i -- \ref{fig:fig-1}c-iii for droplets containing 0.10\% (by mass) and 0.50\% (by mass) PEO1M, respectively, reveals its salient features (see movies SM1 and SM2 in the Supplementary Information for the corresponding movies). The process initiates with the formation of radial interfacial perturbations along the droplet perimeter, characterized by a distinct wavelength. As the droplet continues to spread, these perturbations evolve into finger-like structures that extend outward from the contact line. Eventually, the fingers undergo capillary-driven fragmentation, continuously shedding daughter droplets from the rim until the parent droplet is fully emulsified. \\

While the overall morphology of self-emulsification---rim destabilization, fingering, and fragmentation---remains qualitatively consistent across the polymer concentration, $C_{\mathrm{m}}$, range studied, the dynamics of the process are strongly modulated by $C_{\mathrm{m}}$. In the sections that follow, we quantitatively analyze how polymer-induced viscoelasticity influences key features such as bursting onset time, instability wavelength, finger length, and size distribution of daughter droplets. 

\subsection{Viscoelasticity enhances droplet lifetime and delays self-fragmentation}

\begin{figure}
    \centering
    \includegraphics[width=\textwidth]{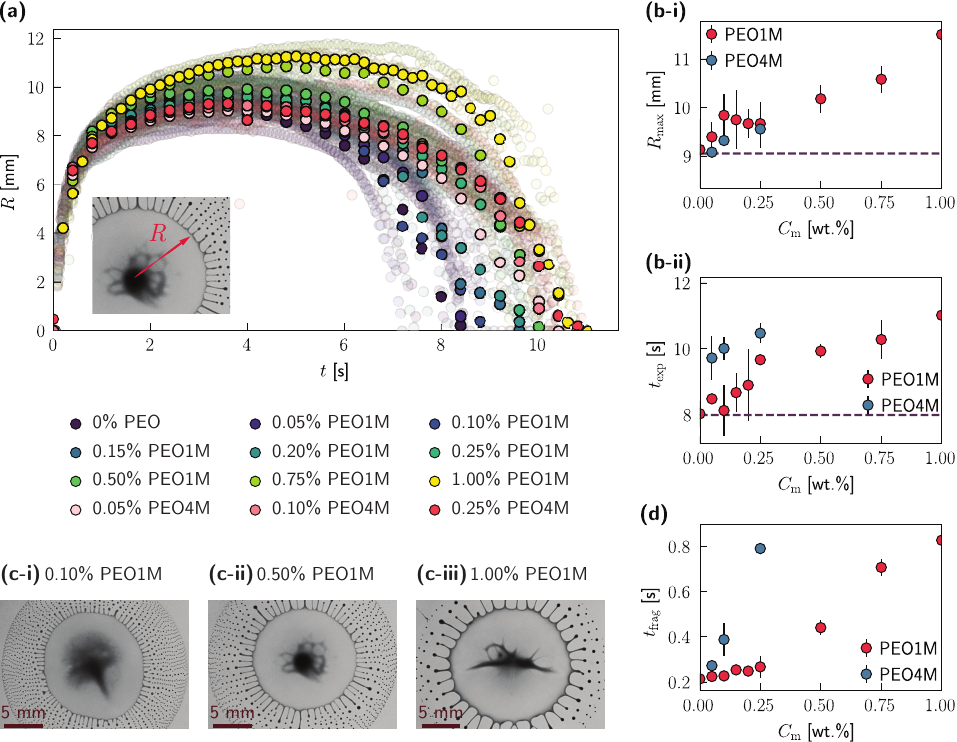}
    \caption{\textbf{Spreading dynamics.} \textbf{(a)} Temporal variation of the spreading front radius, $R$, for water-IPA droplets for different concentrations of the dissolved polymer, where the translucent symbols denote at least three independent experimental realizations per polymer concentration while the opaque symbols indicate the mean for each polymer concentration; the inset shows a typical measurement of $R$ from experimental snapshots. See movie SM3 in the Supplementary Information for the corresponding movie. \textbf{(b-i)} $R_{\max}$ and \textbf{(b-ii)} $t_{\exp}$ as a function of the polymer concentration, $C_{\mathrm{m}}$; the dashed lines indicate the prediction for Newtonian liquids \cite{keiser-2017-prl} (see equations \eqref{eq:keiser-a} and \eqref{eq:keiser-b}). \textbf{(c-i) -- (c-iii)} Snapshots at $t = 0.5 \, t_{\exp}$ for water-IPA droplets for different polymer concentrations. \textbf{(d)} Time of incipience of self-fragmentation, $t_{\mathrm{frag}}$, for water-IPA droplets as a function of polymer concentration, $C_{\mathrm{m}}$. In panels b-i, b-ii, and d, the discrete markers denote the mean of at least three independent experimental realizations while the error bars indicate $\pm$ one standard deviation.}
    \label{fig:fig-2}
\end{figure}

To quantify the influence of viscoelasticity on the dynamics of Marangoni bursting, we track the temporal evolution of the spreading front radius, $R(t)$ (as depicted in the inset of figure \ref{fig:fig-2}a), which captures both the spreading and fragmentation phases of the droplet. The influence of the polymer concentration, $C_{\mathrm{m}}$, on the spreading dynamics is shown in figure \ref{fig:fig-2}a (see movie SM3 in the Supplementary Information for a typical spreading dynamics). \\

For all the liquids tested in the present work, the temporal variation of the spreading front radius, $R$, exhibits three distinct regimes (see figure \ref{fig:fig-2}a): an initial rapid spreading regime ($t \approx$  0 -- 2 s), a quasi-steady plateau regime where the droplet reaches a maximal spreading radius, $R_{\max}$ ($t \approx$ 2 -- 7 s), and a final receding regime ($t \gtrapprox$ 7 s) where $R$ decreases with $t$ while daughter droplets are continually ejected from the periphery of the mother droplet. The spreading dynamics concludes when the entire mother droplet has self-fragmented into daughter droplets, marked by $R \approx 0$ at time $t = t_{\exp}$. Interestingly, while the initial rapid spreading is independent of $C_{\mathrm{m}}$ (consistent with observations on the spreading of polymeric droplets on solid substrates \cite{bartolo-2007-prl, gorin-2022-langmuir, sen-2022-arxiv}; see also figure \ref{fig:fig-s3}a in the Supplementary Information), both $R_{\max}$ and $t_{\exp}$ increases with $C_{\mathrm{m}}$, as shown in figures \ref{fig:fig-2}b-i and \ref{fig:fig-2}b-ii. This suggests that viscoelasticity significantly extends the spatial reach and lifetime of the spreading droplet before rupture. Similar trends are also observed when increasing the initial droplet volume, consistent with previous findings for Newtonian droplets \cite{keiser-2017-prl}. \\ 

To illustrate this effect, figures \ref{fig:fig-2}c-i -- \ref{fig:fig-2}c-iii show side-by-side snapshots at $t = 0.5 t_{\exp}$ for three droplets with increasing PEO1M concentrations. Although each droplet is at the same normalized lifetime, the extent of fragmentation varies widely: the lowest concentration ($C_{\mathrm{m}}$ = 0.10\%, figure \ref{fig:fig-2}c-i) has already released a large number of daughter droplets, while the highest concentration ($C_{\mathrm{m}}$ = 1.00\%, figure \ref{fig:fig-2}c-iii) shows minimal fragmentation and a more extended parent droplet. These results reinforce the idea that higher viscoelasticity delays the onset of fragmentation, allowing the mother droplet to sustain deformation for a longer time and over a larger area. \\

When the spreading dynamics are rescaled using normalized coordinates ($R/R_{\max}$ and $t/t_{\exp}$), the data collapes onto a single master curve (see figure \ref{fig:fig-s3}b in the Supplementary Information) for all $C_{\mathrm{m}}$, consistent with prior observations in Newtonian systems \cite{keiser-2017-prl}. Moreover, the characteristic scaling laws previously proposed \cite{keiser-2017-prl} for Newtonian droplets predict that $R_{\max} \sim R^{\ast}$ and $t_{\exp} \sim t^{\ast}$, where $R^{\ast}$ and $t^{\ast}$ are the characteristic length and time scales, respectively, given by 
\begin{subequations}
    \begin{align}
        R^{\ast} &\sim \left( \frac{ \left( \phi_{0} - \phi_{\mathrm{c}} \right) \Delta \gamma \, h_{\mathrm{o}} \Omega_{0} }{ \left( 1 - \phi_{\mathrm{c}} \right) \eta_{\mathrm{o}} j_{\mathrm{v}}} \right)^{1/4} \, , \label{eq:keiser-a} \\
        t^{\ast} &\sim \left( \frac{ \left( \phi_{0} - \phi_{\mathrm{c}} \right) \eta_{\mathrm{o}} \Omega_{0} }{ \left( 1 - \phi_{\mathrm{c}} \right) \Delta \gamma \, h_{\mathrm{o}} j_{\mathrm{v}}} \right)^{1/2} \, , \label{eq:keiser-b}
    \end{align}
\end{subequations}
where $\phi_{0}$ and $\phi_{\mathrm{c}}$ are, respectively, the initial and critical IPA concentrations, $\Omega_{0}$ the initial mother droplet volume, $\eta_{\mathrm{o}}$ and $h_{\mathrm{o}}$, respectively, the viscosity and depth of the oil bath, $\Delta \gamma$ the interfacial tension difference driving the Marangoni flow, and $j_{\mathrm{v}}$ the evaporation rate of IPA (see the Methods section for a detailed derivation of equations \eqref{eq:keiser-a} and \eqref{eq:keiser-b}). The dashed lines in figures \ref{fig:fig-2}b-i and \ref{fig:fig-2}b-ii describe these proposed scaling relations (as $R_{\max} = 0.28 R^{\ast}$ and $t_{\exp} = 1.55 t^{\ast}$, where the prefactors are determined by fitting equations \eqref{eq:keiser-a} and \eqref{eq:keiser-b} to the droplet without any polymers, i.e. $C_{\mathrm{m}} = 0$, and are close to the values used in previous studies \cite{keiser-2017-prl}). The parameters considered in equations \eqref{eq:keiser-a} and \eqref{eq:keiser-b} remain fairly independent of the $C_{\mathrm{m}}$ range in the present work (see the Supplementary Information for further details). Yet, significant deviations from the proposed scalings appear at increasing $C_{\mathrm{m}}$ (see figures \ref{fig:fig-2}b-i and \ref{fig:fig-2}b-ii), indicating that polymer-induced viscoelastic effects are not captured by existing Newtonian models. Notably, these deviations are even more pronounced for the higher molecular weight polymer (PEO4M), especially at low concentrations---highlighting the sensitivity of the bursting dynamics to the molecular properties of the polymer. \\

Another key descriptor of the bursting behavior is the onset time for fragmentation, $t_{\mathrm{frag}}$, defined as the moment when the first daughter droplets are visibly ejected. As shown in figure \ref{fig:fig-2}d, $t_{\mathrm{frag}}$ increases with $C_{\mathrm{m}}$, further confirming that viscoelasticity delays the initiation of self-emulsification (see also figure \ref{fig:fig-s3}c in the Supplementary Information). For instance, a droplet with a lower $C_{\mathrm{m}}$ (e.g. 0.10\% PEO1M) begins to fragment almost instantaneously ($t_{\mathrm{frag}} \approx$ 0.23 s), while a larger $C_{\mathrm{m}}$ droplet (e.g. 1.00\% PEO1M) exhibits delayed ejection($t_{\mathrm{frag}} \approx$ 0.83 s). This extended onset correlates with the higher values of $R_{\max}$ and $t_{\exp}$ (as reported in figures \ref{fig:fig-2}b-i and \ref{fig:fig-2}b-ii), suggesting that viscoelasticity plays a critical role in regulating both the timing and spatial extent of the fragmentation cascade. \\

Together, these findings establish that polymer concentration---and by extension, viscoelasticity---can serve as a tunable control parameter to program the lifetime, maximum extent, and fragmentation onset of droplets undergoing Marangoni bursting. This tunability offers a route to precisely engineer emulsification dynamics in interfacial material systems. 

\subsection{Viscoelasticity increases the wavelength of azimuthal interfacial instabilities}

\begin{figure}
    \centering
    \includegraphics[width=\textwidth]{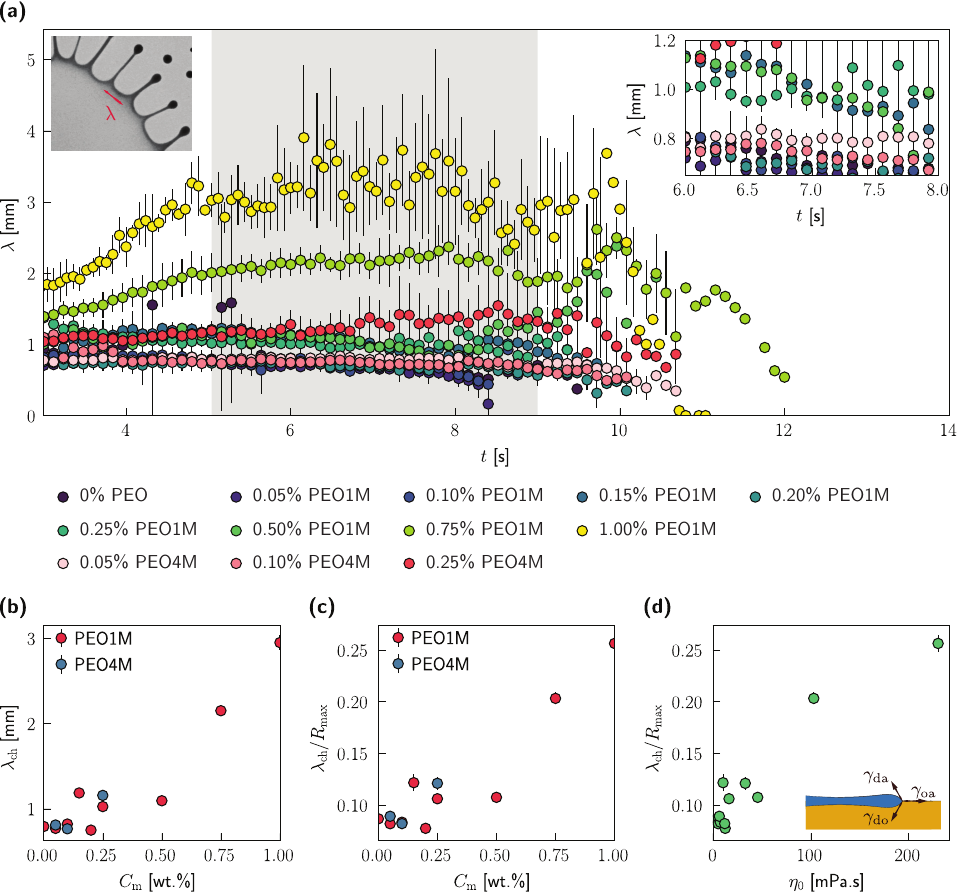}
    \caption{\textbf{Instability wavelength.} \textbf{(a)} Temporal variation of the instability wavelength, $\lambda$, for different polymer concentrations. The shaded area denotes the time-span corresponding to a quasi-constant $\lambda$. The left inset shows a typical measurement of $\lambda$ from experimental snapshots while the right inset shows a zoomed-in comparison of the temporal variation of $\lambda$ for different polymer concentrations. \textbf{(b)} Variation of characteristic wavelength, $\lambda_{\mathrm{ch}}$, with polymer concentration, $C_{\mathrm{m}}$. Normalized characteristic wavelength, $\lambda_{\mathrm{ch}} / R_{\max}$, expressed as a function of \textbf{(c)} the polymer concentration, $C_{\mathrm{m}}$, and \textbf{(d)} the  zero-shear viscosity, $\eta_{0}$. The inset of panel d shows the balance of interfacial tensions at the drop-oil-air interface. In each panel, the discrete markers denote the mean of at least three independent experimental realizations while the error bars indicate $\pm$ one standard deviation.}
    \label{fig:fig-3}
\end{figure}

The self-emulsification process in Marangoni bursting is initiated via a destabilization of the liquid rim at the perimeter of the spreading droplet, as shown in figures \ref{fig:fig-1}b-i -- \ref{fig:fig-1}b-iii and \ref{fig:fig-1}c-i -- \ref{fig:fig-1}c-iii. This instability manifests as an azimuthal modulation of the droplet spreading front, forming periodic finger-like protrusions that later fragment into daughter droplets. The spatial periodicity of these perturbations is characterized by a wavelength, $\lambda$, as illustrated in the left inset of figure \ref{fig:fig-3}a. \\

Due to the inherent complexity and transient nature of the bursting dynamics, the number and spacing of the fingers can fluctuate significantly during each experiment (see figures \ref{fig:fig-1}b-i -- \ref{fig:fig-1}b-iii, \ref{fig:fig-1}c-i -- \ref{fig:fig-1}c-iii, and movies SM1 and SM2 in the Supplementary Information). To estimate a representative wavelength, we adopt an alternate approach \cite{seyfert-2022-prf}, defining the instantaneous wavelength as $\lambda(t) = 2 \pi R(t) / n_{\mathrm{finger}} (t)$, where $n_{\mathrm{finger}}$ denotes the number of protruding fingers from the spreading front, located at a radial location $R$, at time $t$. This approximation is valid for cases where the droplet circumference is much greater than the characteristic wavelength (i.e. $2 \pi R \gg \lambda$). The temporal evolution of $\lambda$ for different $C_{\mathrm{m}}$ values is shown in figure \ref{fig:fig-3}a (the corresponding temporal variation of $n_{\mathrm{finger}}$ is shown in figure \ref{fig:fig-s4} in the Supplementary Information). Note that our simplified approach results in $\lambda$ measurements within the same numerical range as previously reported measurements \cite{keiser-2017-prl}. We present measurements for $t \gtrapprox$ 3 s due to the underestimation of $n_{\mathrm{finger}}$ at early times (see the Supplementary Information for further details). \\ 

For all polymer concentrations, the instability wavelength exhibits a similar temporal evolution: a brief initial increase followed by a quasi-plateau phase (marked by the shaded region in figure \ref{fig:fig-3}a), and finally a decrease as fragmentation proceeds. Notably, the quasi-plateau phase of $\lambda$ aligns temporally with the quasi-plateau in the spreading dynamics (figure \ref{fig:fig-2}a) close to the maximum spreading radius, $R_{\max}$, of the droplet. Importantly, the quasi-plateau value of $\lambda$ increases systematically with polymer concentration, suggesting that viscoelasticity plays a stabilizing role in suppressing short-wavelength instabilities. \\

To quantify this trend, we define a characteristic wavelength \cite{seyfert-2022-prf}, $\lambda_{\mathrm{ch}}$, as the mean $\lambda$ within the temporal window $t$ = $0.5 t_{\exp}$ $\pm$ 0.2 s, where the azimuthal features are well-developed. As shown in figure \ref{fig:fig-3}b, $\lambda_{\mathrm{ch}}$ remains virtually invariant with $C_{\mathrm{m}}$ at $\approx$ 1 mm for $C_{\mathrm{m}}$ $<$ 0.50\%, consistent with previous measurements for Newtonian droplets \cite{keiser-2017-prl, hasegawa-2021-pof}. However, at higher polymer concentrations, $\lambda_{\mathrm{ch}}$ increases significantly, reaching values as high as $\approx$ 3 mm for $C_{\mathrm{m}}$ = 1.00\% PEO1M, demonstrating a clear polymer concentration-dependent suppression of high-frequency (short-wavelength) interfacial instabilities. \\

From a mechanistic perspective, the instability wavelength in Marangoni bursting is known to depend on the interfacial tension gradient, $\Delta \gamma / R$ (see the Methods section for further details), which acts as the driving force for flow instabilities \cite{keiser-2017-prl, seyfert-2022-prf}: a stronger gradient is known to result in a shorter wavelength and vice versa. In our experiments, the parameters governing $\Delta \gamma$ (e.g. the initial concentration of IPA \cite{keiser-2017-prl}, the concentration of colorants such as Methylene Blue \cite{seyfert-2022-prf}) are held constant (see the Methods section for further details). Hence, any change in the interfacial tension gradient arises primarily from differences in the radial extent $R$, which increases with polymer concentration (as shown in figure \ref{fig:fig-2}). Therefore, an increase in $C_{\mathrm{m}}$ leads to a reduction in the magnitude of $\Delta \gamma / R$,  partially explaining the observed increase in $\lambda_{\mathrm{ch}}$. \\

To compensate for this effect, we normalize $\lambda_{\mathrm{ch}}$ with $R_{\max}$, and still observe a monotonic increase in $\lambda_{\mathrm{ch}} / R_{\max}$ with $C_{\mathrm{m}}$ (figure \ref{fig:fig-3}c). This result suggests that the interfacial tension gradient alone does not fully account for the wavelength selection mechanism. \\

Indeed, additional insights are revealed in in figure \ref{fig:fig-3}d, where $\lambda_{\mathrm{ch}} / R_{\max}$ is plotted against the zero-shear viscosity, $\eta_{0}$, of the polymer solutions. A positive correlation emerges, indicating that fluid rheology---particularly viscous resistance to interface deformation---also contributes to the stabilization of the long-wavelength modes. Although a full (non-)linear stability analysis is beyond the scope of the present work, these findings point to a multifactorial dependence of the azimuthal instability on both interfacial and rheological properties of the spreading droplet. \\

Collectively, the findings described above establishes that viscoelasticity increases the dominant wavelength of azimuthal interfacial instabilities. This tunability is of particular interest for applications requiring controlled emulsification, pattern formation, or dynamic interface engineering in soft materials.  

\subsection{Polymer-induced viscoelasticity promotes longer fingers prior to fragmentation}

\begin{figure}
    \centering
    \includegraphics[width=\textwidth]{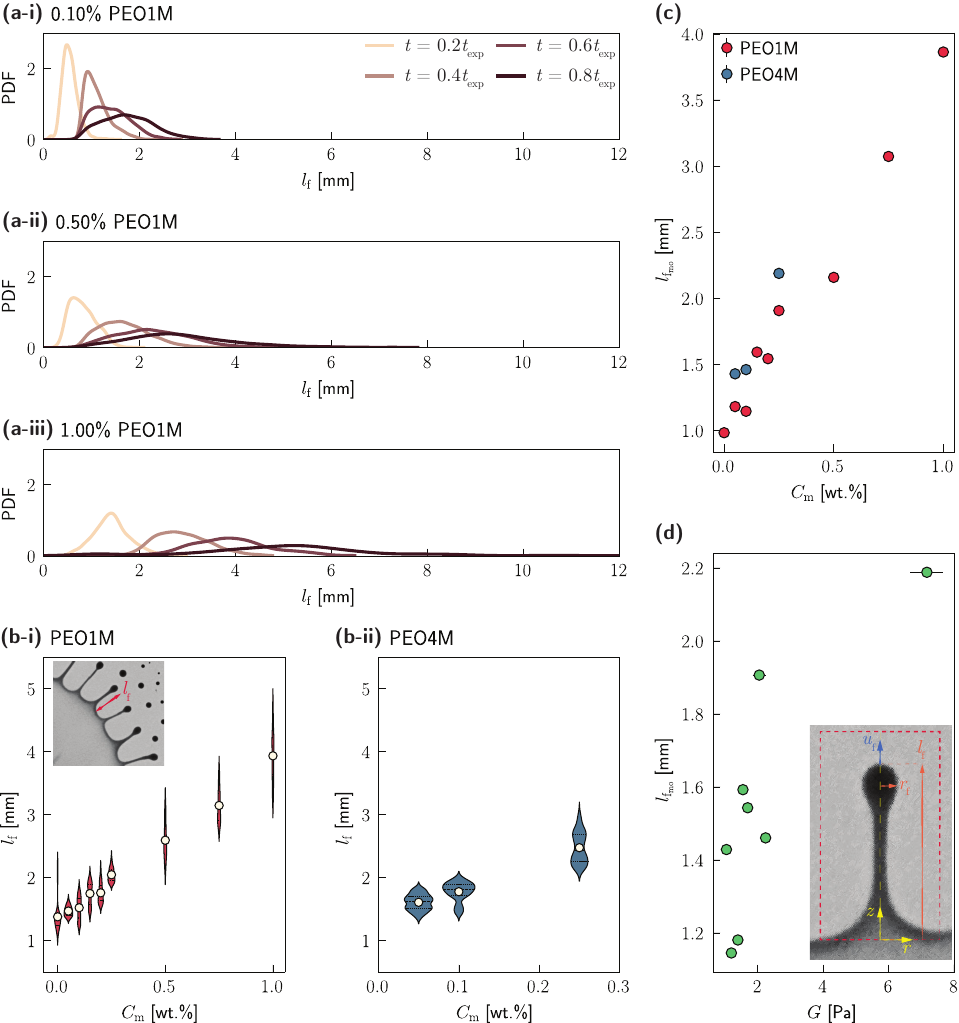}
    \caption{\textbf{Finger length.} \textbf{(a-i) -- (a-iii)} Distributions of the finger length, $l_{\mathrm{f}}$, for different polymer concentrations, where the different colors indicate different time instants. Finger length, $l_{\mathrm{f}}$, distributions at $t = 0.5 \, t_{\exp}$ for different concentrations, $C_{\mathrm{m}}$, of \textbf{(b-i)} PEO1M and \textbf{(b-ii)} PEO4M, where the discrete datapoints denote the mean value of each distribution. The inset in panel b-i shows a typical measurement of $l_{\mathrm{f}}$ from experimental snapshots. Variation of the characteristic finger length, $l_{\mathrm{f}_{\mathrm{mo}}}$, at $t = 0.6 t_{\exp}$ with \textbf{(c)} the polymer concentration, $C_{\mathrm{m}}$, and \textbf{(d)} the elastic modulus, $G$; the discrete symbols denote the mean of at least three independent measurements while the error bars indicate $\pm$ one standard deviation. The inset in panel d shows the schematic of the theoretical model for finger stretching, also clarifying the employed notation.}
    \label{fig:fig-4}
\end{figure}

Following the onset of the azimuthal interfacial instability, finger-like protrusions emerge from the periphery of the spreading droplet (as shown in figure \ref{fig:fig-1}), which grow in length with time and ultimately fragment into smaller daughter droplets. We examine the evolution of these fingers by measuring their length, $l_{\mathrm{f}}$, at the moment just prior to breakup (a typical measurement is illustrated in the inset of figure \ref{fig:fig-4}b-i). Distributions of $l_{\mathrm{f}}$ at various normalized lifetimes, $t/t_{\exp}$, are shown for three representative polymeric concentrations, $C_{\mathrm{m}}$, in figures \ref{fig:fig-4}a-i -- \ref{fig:fig-4}a-iii. \\ 

Across all conditions, the finger length increases with time, as seen from both the broadening and the rightward shift of the distributions in figures \ref{fig:fig-4}a-i -- \ref{fig:fig-4}a-iii. More importantly, increasing polymer concentration leads to significantly longer and more broadly distributed fingers at any time point, as evidenced by figures \ref{fig:fig-4}b-i and \ref{fig:fig-4}b-ii. This suggests that viscoelasticity---imparted by polymer additives----enables the fingers to sustain elongation (or stretching) for longer periods before capillary breakup occurs. \\

To quantify this trend, we define a characteristic finger length, $l_{\mathrm{f_{\mathrm{mo}}}}$, as the modal value of the finger length distribution at each time instant. Since $l_{\mathrm{f}}$ is defined as the length of the finger just prior to the pinch-off of daughter droplets, $l_{\mathrm{f_{\mathrm{mo}}}}$ denotes, for each $C_{\mathrm{m}}$, the typical length to which fingers can be stretched before they eventually break up. The variation of $l_{\mathrm{f_{\mathrm{mo}}}}$ with $C_{\mathrm{m}}$, shown in figure \ref{fig:fig-4}c for $t = 0.6 t_{\exp}$, demonstrates a monotonic increase. Note that while the precise numerical values of $l_{\mathrm{f}_{\mathrm{mo}}}$ are different at different time instants, the trend with changing $C_{\mathrm{m}}$ remains qualitatively the same as the one shown in figure \ref{fig:fig-4}c for $t = 0.6 t_{\exp}$. Remarkably, droplets with 1.00\% PEO1M form fingers that are nearly 300\% longer than those from Newtonian (polymer-free, i.e. $C_{\mathrm{m}}$ = 0\%) systems. Furthermore, increasing the molecular weight of the polymeric additive (from PEO1M to PEO4M) at a given concentration also results in longer fingers, thus reinforcing the role of fluid elasticity in governing this behavior. \\

To quantitatively interpret this viscoelastic stretching mechanism, we develop a simplified theoretical model to capture the dominant force balance within a stretching finger. In a control volume containing a stretching finger that is always bounded by the inflection points at the drop-oil interface, as shown by the dashed rectangle in the inset of figure \ref{fig:fig-4}d, an axisymmetric ($r$-$z$) coordinate system, co-moving with the periphery (or rim) of the spreading droplet is considered. Given the small radial-to-axial length scale of the fingers, we formalize the stretching dynamics using the slender jet approximation \cite{eggers-1993-prl, eggers-1994-jfm, shi-1994-nature, clasen-2006-jfm, book-eggers, sen-2021-jfm, dixit-2024-arxiv}, within which the axial momentum equation can be written as
\begin{align}
	\rho_{\mathrm{d}} \left( \frac{\partial u}{\partial t} + u \frac{\partial u}{\partial z} \right) = - \gamma_{\mathrm{c}} \frac{\partial \kappa}{\partial z} + \frac{1}{r_{\mathrm{f}}^{2}} \frac{\partial}{\partial z} \left( r_{\mathrm{f}}^{2} \left( 3 \eta_{\mathrm{s}} \frac{\partial u}{\partial z} + G \left( A_{zz} - 1 \right) \right) \right) \, ,
	\label{eq:slender-jet}
\end{align}
where $r_{\mathrm{f}} (z,t)$ and $u (z,t)$, respectively, are the finger radius and axial velocity of the fluid within the stretching finger, $\gamma_{\mathrm{c}}$ the interfacial tension coefficient at the periphery of the mother droplet, $\kappa$ the curvature of the finger, $\eta_{\mathrm{s}}$ the shear viscosity of the solvent phase (aqueous solution of IPA of volume fraction $\phi_{\mathrm{c}}$), and $A_{zz}$ the axial component of the polymer conformation tensor $\mathbb{A}$. In writing equation \eqref{eq:slender-jet}, we have further assumed the Oldroyd-B constitutive relation \cite{oldroyd-1950-prsa, book-bird} for the polymeric stress, which has been successfully used to describe the thinning of viscoelastic liquid filaments \cite{clasen-2006-jfm, eggers-2020-jfm, sen-2021-jfm}. The conformation tensor $\mathbb{A}$ evolves by linear relaxation dynamics in the Oldroyd-B model, where each polymer molecule is pictured as two beads connected by a spring \cite{book-bird}. Integrating over the control volume shown in the inset of figure \ref{fig:fig-4}d, with a differential volume element $\mathrm{d} \Omega = \pi \left( r_{\mathrm{f}}(z, t) \right)^{2} \mathrm{d}z$, allows us to write a force balance given by \cite{trouton-1906-procrsoclond}
\begin{align}
	\frac{\mathrm{d} M_{\mathrm{f}}}{\mathrm{d} t} = 3 \eta_{\mathrm{s}} r_{\mathrm{f}}^{2} \frac{\partial u}{\partial z} \bigg \rvert_{z = 0} + G \, r_{\mathrm{f}}^{2} \left( A_{zz} - 1 \right) \bigg \rvert_{z = 0} \, ,
	\label{eq:momentum-change}
\end{align}
where $M_{\mathrm{f}}(t) = \int_{\Omega (t)} \pi \rho_{\mathrm{d}} \left( r_{\mathrm{f}} (z,t) \right)^{2} u (z,t) \mathrm{d}z$ is the momentum of the stretching finger. The integral of the first term on the right-hand side of equation \eqref{eq:slender-jet} vanishes (and does not appear in equation \eqref{eq:momentum-change}), since the choice of our control volume (see inset of figure \ref{fig:fig-4}d) ensures its orthogonal intersection with the drop-oil interface \cite{marchand-2011-ajp, thesis-munro, dixit-2024-arxiv}. Additionally, the integral of the second term on the right-hand side of equation \eqref{eq:slender-jet} vanishes at $z = l_{\mathrm{f}}(t)$ since $r_{\mathrm{f}} (z = l_{\mathrm{f}}(t), t) = 0$ at the tip of the stretching finger. Now, the first term on the right-hand side of equation \eqref{eq:momentum-change}, arising due to the viscosity of the solvent phase (water-IPA mixture of volume fraction $\phi_{\mathrm{c}}$), has a near-insignificant contribution to the stretching dynamics since $\eta_{\mathrm{s}}$ has a small contribution to the overall shear viscosity $\eta$ (see figure \ref{fig:fig-s2}a in the Supplementary Information), and is independent of the polymer concentration $C_{\mathrm{m}}$. Hence, we can deduce from equation \eqref{eq:momentum-change} that
\begin{align}
	\frac{\mathrm{d} M_{\mathrm{f}}}{\mathrm{d} t} \sim G \, r_{\mathrm{f}}^{2} \left( A_{zz} - 1 \right) \bigg \rvert_{z = 0} \, .
	\label{eq:momentum-sim-1}
\end{align}
If $u_{\mathrm{f}}$ is the characteristic stretching velocity of the fingers, equation \eqref{eq:momentum-sim-1} can be recast as 
\begin{align}
	\rho_{\mathrm{d}} u_{\mathrm{f}}^{2} r_{\mathrm{f}}^{2} \sim G \, r_{\mathrm{f}}^{2} \left( A_{zz} - 1 \right) \bigg \rvert_{z=0} \, ,
	\label{eq:momentum-sim-2}
\end{align}
which leads to
\begin{align}
	u_{\mathrm{f}} \sim \left( \frac{G}{\rho_{\mathrm{d}}} \left( A_{zz} - 1 \right) \bigg \rvert_{z=0} \right)^{1/2} \, .
	\label{eq:velocity-scale}
\end{align}
Now, the conformation tensor $\mathbb{A}$ is related to the individual polymer molecules within the liquid as \cite{book-bird} $\mathbb{A} = \langle \mathbf{X} \mathbf{X} \rangle / X_{\mathrm{eq}}^{2}$, where each polymer molecule is stretched to a length $\mathbf{X}$ from its equilibrium length $X_{\mathrm{eq}}$. Hence, $\left( A_{zz} - 1 \right) \rvert_{z=0} ^{1/2}$ in equation \eqref{eq:velocity-scale} is linearly related to the local polymer stretching \cite{sen-2021-jfm}. However, quantifying the microscale polymer stretching dynamics from macroscale, continuum-level experiments, such as the ones described in the present work, is an arduous task. Additionally, a limitation of the Oldroyd-B model is that it assumes the polymers to be infinitely extensible while, in reality, the dissolved polymers have a finite extensibility limit. This finite extensibility becomes important especially when an axially-thinning liquid filament (e.g. the stretching liquid fingers in the inset of figure \ref{fig:fig-4}b-i) breaks up to produce daughter droplets. The experimental determination of the finite extensibility limit of polymers is a challenge. Moreover, incorporating this finite extensibility into the analysis also necessitates a nonlinear constitutive relation, which comes with additional (unknown) fitting parameters \cite{sen-2021-jfm, sen-2022-arxiv}. Furthermore, other factors may also play a role in the stretching dynamics, such as polydispersity and multiple relaxation time scales of the polymer molecular chains \cite{entov-1997-jnnfm, wagner-2005-prl}. These limitations, unfortunately, prevent a one-to-one comparison between the experimental results and the theoretical model. Nevertheless, the strength of this theoretical model lies in a quantitative, physically grounded prediction of how the finger stretching velocity, $u_{\mathrm{f}}$, scales with the elastic modulus, $G$, of the polymeric liquid (see equation \eqref{eq:velocity-scale}). Consequently, the characteristic achievable finger length, $l_{\mathrm{f}_{\mathrm{mo}}}$, is expected to increase with increasing $G$, and thus increasing polymer concentration, $C_{\mathrm{m}}$ -- consistent with our experiments (as demonstrated in figures \ref{fig:fig-4}d and \ref{fig:fig-4}c, respectively). \\

In summary, both experimental results and theoretical analysis converge on the conclusion that polymer-induced viscoelasticity enables significantly longer finger growth prior to droplet breakup. This presents a powerful mechanism to tune fragmentation length scales in Marangoni bursting, and more broadly, to design soft fluidic systems where the breakup dynamics can be predictively controlled through molecular-level modifications of fluid rheology. 

\subsection{Viscoelasticity modulates emulsification timing, not droplet size}

\begin{figure}
    \centering
    \includegraphics[width=\textwidth]{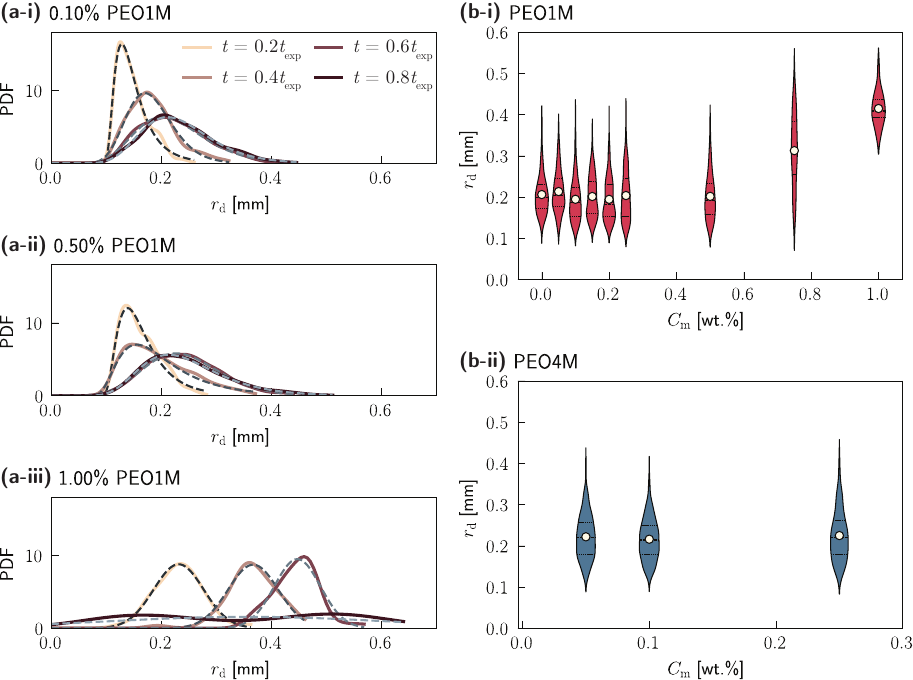}
    \caption{\textbf{Size of daughter droplets.} \textbf{(a-i) -- (a-iii)} Distributions of the daughter droplet radius, $r_{\mathrm{d}}$, for different polymer concentrations, where the different colors indicate different time instants. The dashed lines denote the corresponding log-normal distribution fits. Daughter droplet radius, $r_{\mathrm{d}}$, distributions at $t = 0.5 \, t_{\exp}$ for different concentrations, $C_{\mathrm{m}}$, of \textbf{(b-i)} PEO1M and \textbf{(b-ii)} PEO4M, where the discrete datapoints denote the mean value of each distribution. }
    \label{fig:fig-5}
\end{figure}

During the Marangoni bursting process, radial fingers extending from the spreading droplet stretch and eventually fragment into a large population of daughter droplets (as shown in figure \ref{fig:fig-1}). We quantify the outcome of this fragmentation by analyzing the droplet size distribution, specifically the radial size, $r_{\mathrm{d}}$, for different polymer concentrations, $C_{\mathrm{m}}$. The temporal evolution of $r_{\mathrm{d}}$ distributions for three representative concentrations of PEO1M is presented in figures \ref{fig:fig-5}a-i -- \ref{fig:fig-5}a-iii. In each case, experimental data (solid lines) are well-fitted by log-normal distributions (dashed lines), consistent with established theories of fragmentation \cite{villermaux-2007-arfm}.  \\

For both $C_{\mathrm{m}}$ = 0.10\% and 0.50\% PEO1M (figures \ref{fig:fig-5}a-i and \ref{fig:fig-5}a-ii), we observe that the droplet size distribution broadens over time, accompanied by a mild shift toward larger droplet radii. At a higher polymer concentration ($C_{\mathrm{m}}$ = 1.00\%, figure \ref{fig:fig-5}a-iii), the distribution does not broaden, but instead shifts consistently toward larger droplet sizes with time. Additionally, at late stages (e.g. $t = 0.8 \, t_{\exp}$), the total number of detectable daughter droplets significantly decreases for the highest $C_{\mathrm{m}}$ due to nearly complete fragmentation of the parent droplet, as well as evaporative shrinkage caused by the volatile IPA content in the daughter droplets \cite{seyfert-2022-prf}. This effect is amplified at higher $C_{\mathrm{m}}$ values, since droplet lifetimes increase with polymer concentration (figure \ref{fig:fig-2}b-ii), thus providing a longer time window for evaporation to occur. \\ 

To directly assess the effect of viscoelasticity on the final daughter droplet size, we compare size distributions at a fixed normalized time $t = 0.5 \, t_{\exp}$ across various $C_{\mathrm{m}}$ for both PEO1M and PEO4M (see figures \ref{fig:fig-5}b-i and \ref{fig:fig-5}b-ii). For $C_{\mathrm{m}}$ $\leq$ 0.50\%, the mean daughter droplet size remains largely unchanged, regardless of polymer concentration or molecular weight. At higher concentrations, a modest increase in droplet size is observed for PEO1M, consistent with prior reports on droplet formation in viscoelastic jets \cite{christanti-2001-jnnfm, brenn-2017-atsprays}. However, these trends should be interpreted with caution, as the sample size at high $C_{\mathrm{m}}$ is limited due to reduced droplet counts (resulting from a reduced number of fingers; see figure \ref{fig:fig-s4} in the Supplementary Information) and evaporative shrinkage. \\

Notably, even if the mean droplet size remains constant, the shape of the size distribution is sensitive to viscoelasticity. With increasing $C_{\mathrm{m}}$, the distributions become broader, suggesting that polymer-induced viscoelasticity may impact the uniformity and breakup dynamics, even if the final droplet dimensions are statistically similar. \\

More significantly, viscoelasticity has a pronounced effect on the timing (or onset) of self-emulsification. As shown in figure \ref{fig:fig-2}d (see also figure \ref{fig:fig-s3}c in the Supplementary Information), the onset time of fragmentation, $t_{\mathrm{frag}}$, increases with polymer concentration. In other words, while the size of the resulting droplets may not change dramatically with $C_{\mathrm{m}}$, the moment at which these droplets form can be precisely delayed by tuning the viscoelastic properties of the constituent fluid. Additionally, we also noticed an upper bound for polymer concentration, where the viscous and viscoelastic effects are so strong that self-emulsification of the droplets is completely arrested (see movie SM4 in the Supplementary Information). \\

Taken together, these results highlight a non-intuitive design principle: polymer-induced viscoelasticity does not significantly influence the final droplet size, but acts as a temporal control mechanism that modulates the onset and progression of self-emulsification. This insight is critical for applications where the timing of fragmentation, rather than the droplet dimensions alone, governs performance---such as in triggered release systems, programmable emulsions, or responsive interfacial materials. 

\section{Conclusions and outlook}

In conclusion, we have established a materials-based strategy for tuning interfacial instabilities by introducing polymer-induced viscoelasticity into Marangoni-driven self-emulsification processes. By carefully controlling the polymer concentration, we demonstrate the ability to modulate key features of the instability, including maximum spreading, droplet lifetime, and the wavelength of the emerging patterns. Viscoelasticity not only stabilizes the interfacial dynamics---delaying the onset of self-emulsification---but also enables significantly longer finger stretching before breakup, resulting in fewer, more widely-spaced daughter droplets. These effects are captured quantitatively by a scaling law linking finger stretching dynamics to fluid elasticity, offering a predictive framework for the design of viscoelastic interfaces with a surface tension gradient. \\

Together with high-resolution experimental observations, our findings bridge fundamental fluid dynamics with interfacial material design. This work introduces a controllable, surface-tension-mediated mechanism to engineer interfacial behavior, opening pathways for responsive emulsions, programmable droplet generation, and microfluidic appications that demand precise control over interfacial transport and breakup. 

\section{Acknowledgements}

The authors are grateful for the technical assistance from Remco Fokkink and Raoul Fix during the fabrication of the experimental setup. The authors thank Valeria Garbin, Jasper van der Gucht, Steffen Hardt, Maziyar Jalaal, Thomas Kodger, Frans Leermakers, Detlef Lohse, Alvaro Marin, Gareth McKinley, Manikuntala Mukhopadhyay, Vatsal Sanjay, and Jacco Snoeijer for insightful discussions. 

\section{Author contributions}

C.H. and U.S. designed the experimental setup. C.H. carried out the experiments and data analysis. U.S. directed the project, and supervised the experiments and data analysis. All authors contributed to interpreting the experimental observations and writing the manuscript. 

\section{Competing interests}

The authors declare no competing interests.

\section{Data availability}

Data is available from the corresponding author upon request. 

\section{Methods}

\subsection{Preparation of polymeric droplets and substrate}

The droplet solvent phase consisted of a 40\% (by mass) solution of isopropyl alcohol (2-propanol, Thermo-Scientific, henceforth referred to as ``IPA") in purified water (Milli-Q). This IPA concentration exceeds the minimum (critical) alcohol content required to trigger Marangoni bursting, as previously established \cite{keiser-2017-prl}. To impart viscoelasticity, polyethylene oxide (average molecular weights $\approx$ 1 $\times$ 10$^\text{6}$ Da and 4 $\times$ 10$^\text{6}$ Da, Sigma-Aldrich, henceforth referred to as ``PEO1M" and ``PEO4M", respectively), was dissolved into the IPA-water mixture at concentrations (by mass), $C_{\mathrm{m}}$, ranging from 0.05\% to 1.00\%. \\

Following previous protocols \cite{seyfert-2022-prf}, all water-IPA-polymer solutions were additionally dyed with Methylene Blue (Sigma-Aldrich) at a fixed concentration of 0.7 mg/mL to increase optical contrast and enable reliable edge detection during automated image analysis. \\ 

The substrate phase consisted of commercially available sunflower oil (Vandemoorte Nederland BV), sourced from a local supermarket, and used without further purification. 

\subsection{Experimental protocol}

A schematic of the experimental setup is shown in figure \ref{fig:fig-1}a. A polypropylene Petri dish (100 mm diameter, VWR) was filled with sunflower oil to a depth of 5 mm, forming the liquid substrate or ``oil bath". For back-illumination, the Petri dish was then placed atop an LED light pad (L4S LED light pad, Huion, not shown in figure \ref{fig:fig-1}a), which ensured uniform contrast for high-quality imaging. \\

Droplets of the water-IPA-polymer solution (volume $\approx$ 7.5 \SI{}{\micro \text{L}}) were gently deposited onto the oil surface using a disposable syringe (5 mL capacity, Sigma-Aldrich) fitted with a blunt stainless steel precision dispensing tip (inner diameter = 0.41 mm, Nordson EFD). Upon contact, the droplet initiated spontaneous Marangoni bursting. \\

High-resolution optical recordings were captured, at 25 frames-per-second, using a digital mirrorless camera (EOS R6 Mark II, Canon) equipped with a macro objective (RF 35 mm F1.8 IS Macro STM, Canon) and an additional 16 mm lens extension tube (Caruba). This imaging configuration provided a spatial resolution of 12 \SI{}{\micro \text{m}}/pixel over a field of view of 11.5 cm$^{\text{2}}$. Recording began upon droplet contact on the oil bath and continued until the complete fragmentation of the mother droplet into daughter droplets. \\

Each experimental condition was repeated independently at least five times to ensure reproducibility. While the experimental parameters were well-controlled, one-to-one quantitative comparisons between repetitions is limited due to the inherently complex nature of the Marangoni bursting process, as previously reported \cite{keiser-2017-prl, seyfert-2022-prf}. 

\subsection{Image processing and data extraction}

Post-acquisition image analysis was performed using a custom Python script \cite{web-git} based on OpenCV. This pipeline was used to extract key quantitative parameters from each frame, including the spreading front radius, instability wavelength, finger length, and radius of the daughter droplets, corresponding to figures \ref{fig:fig-2} -- \ref{fig:fig-5}. \\

Raw RGB images were first converted to 8-bit grayscale, followed by the application of a median blur to reduce the noise. The images were then binarized using adaptive thresholding, enabling the detection of relevant interfacial features. Binary images were analyzed via contour detection, which privided the foundational geometry for all subsequent measurements. \\

Contours corresponding to fingers were identified by applying filters based on the distance from the droplet center, circularity, and projected area. To estimate the spreading front, the closest points on each contour to the droplet center were isolated, and a circle was fitted through these points using a RANSAC algorithm \cite{fischler-1981-communacm}. The resulting circle provided both the center and the radius of the spreading front. \\

Finger length was calculated by further sorting contours by circularity, distance from the spreading front, area, and orientation. A rotated bounding rectangle was fitted to each identified contour, and the finger length was defined as the longest dimension of the bounding box. \\ 

To extract the radius of the daughter droplets, contours were classified based on their proximity to the spreading front, shape circularity, and area. A droplet-tracking algorithm was implemented to identify newly formed droplets between successive frames, using both the radial distance and angular displacement between detected contours. This allowed for accurate identification and measurement of individual pinch-off events and droplet radii over time.  

\subsection{Scaling analysis of maximum spreading and droplet lifetime}

To estimate the characteristic spreading radius and lifetime of the droplet, we consider the balance of shear stresses: across the liquid-liquid interface between the spreading droplet and the oil substrate, shear stresses must balance each other. This stress balance can be expressed as
\begin{align}
	\sigma_{\mathrm{d}} \sim \sigma_{\mathrm{o}} \, ,
	\label{eq:shear-balance}
\end{align}
where $\sigma_{\mathrm{d}}$ and $\sigma_{\mathrm{o}}$ are the shear stresses in the droplet (subscript "$\mathrm{d}$") and oil (subscript "$\mathrm{o}$") phases, respectively, expressed as
\begin{subequations}
	\begin{align}
		\sigma_{\mathrm{d}} &\sim \eta_{\mathrm{d}} \frac{\Delta u_{\mathrm{d}}}{h_{\mathrm{d}}} \, ,  \label{eq:shear-scaling-a} \\
		\sigma_{\mathrm{o}} &\sim \eta_{\mathrm{o}} \frac{\Delta u_{\mathrm{o}}}{h_{\mathrm{o}}} \, , \label{eq:shear-scaling-b}
	\end{align}
\end{subequations}
where $\eta$ is the shear viscosity and $\Delta u$ the velocity difference across a thickness $h$. It follows from equation \eqref{eq:shear-balance} that
\begin{align}
	\frac{\Delta u_{\mathrm{d}}}{\Delta u_{\mathrm{o}}} \sim \frac{\eta_{\mathrm{o}}/h_{\mathrm{o}}}{\eta_{\mathrm{d}}/h_{\mathrm{d}}} \, .
	\label{eq:vel-ratio}
\end{align}

In our system, $\eta_{\mathrm{o}} / h_{\mathrm{o}} \ll \eta_{\mathrm{d}} / h_{\mathrm{d}}$. Hence, equation \eqref{eq:vel-ratio} implies that $\Delta u_{\mathrm{d}} / \Delta u_{\mathrm{o}} \ll 1$. This suggest that the flow within the spreading droplet can be approximated as a plug flow, allowing us to treat the droplet-oil interface as a  single interface with an effective interfacial tension coefficient $\gamma = \gamma_{\mathrm{da}} + \gamma_{\mathrm{do}}$ (see inset of figure 3d) \cite{keiser-2017-prl}. \\

Now, we can consider the IPA concentration at the center of the mother droplet to be close to the initial concentration $\phi_{0}$, while that at the periphery to be close to the critical concentration, $\phi_{\mathrm{c}}$. Hence, $\gamma$ varies from $\gamma_{0} = \gamma (\phi_{0})$ at the center of the droplet to $\gamma_{\mathrm{c}} = \gamma (\phi_{\mathrm{c}})$ at its periphery. The resulting surface tension gradient $\Delta \gamma / R^{\ast}$ drives a Marangoni flow from the center to the periphery of the mother droplet with a characteristic velocity $u_{\mathrm{d}}$, where $\Delta \gamma = \gamma_{\mathrm{c}} - \gamma_{0}$ and $R^{\ast}$ is the characteristic radius of the mother droplet \cite{keiser-2017-prl}. \\ 

Meanwhile, the flow in the oil phase is initially setup along a boundary layer close to the droplet-oil interface, whose thickness, $\delta$, increases with time $t$ as $\delta \sim \sqrt{\nu t}$, where $\nu$ is the kinematic viscosity of the oil phase. This boundary layer penetrates the entire oil layer thickness $h_{\mathrm{o}}$ in less than a second, which implies that the flow is developed across the entire oil layer for most of the experiment. The viscous stress in the oil layer, $\sigma_{\mathrm{o}}$, must balance the Marangoni stress driving the flow, $\sigma_{\gamma}$, which can be expressed as
\begin{subequations}
	\begin{align}
		\sigma_{\mathrm{o}} &\sim \eta_{\mathrm{o}} \frac{u_{\mathrm{d}}}{h_{\mathrm{o}}} \, , \label{eq:stress-oil} \\
		\sigma_{\gamma} &\sim \frac{\Delta \gamma}{R^{\ast}} \, , \label{eq:stress-marangoni}
	\end{align}
\end{subequations}
resulting in
\begin{align}
	u_{\mathrm{d}} \sim \frac{\Delta \gamma \, h_{\mathrm{o}}}{\eta_{\mathrm{o}} R^{\ast}} \, .
	\label{eq:ud-scale}
\end{align}

From this, we can define the characteristic timescale for the experiments, $t^{\ast}$, to be the timescale for liquid transport from the center of the droplet to its periphery, given by 
\begin{align}
	t^{\ast} \sim \frac{R^{\ast}}{u_{\mathrm{d}}} \, .
	\label{eq:timescale}
\end{align}
Note that this $t^{\ast}$, which is also representative of the typical timescale of spreading, is $\mathcal{O}(1 \, \text{s})$---orders of magnitude larger than the relaxation time, $\tau$, of the polymer-water-IPA solutions used in the present study ($\tau$ is $\mathcal{O}(1 - 10 \, \, \text{ms})$). Hence, during the spreading phase, the dissolved polymers get ample time to initially stretch but eventually relax. Hence, the spreading dynamics are observed to be independent of the polymeric concentration (see figure \ref{fig:fig-2}a and figure \ref{fig:fig-s3}a in the Supplementary Information), which is also consistent with prior observations on the spreading of polymeric droplets on solid substrates \cite{bartolo-2007-prl, gorin-2022-langmuir, sen-2022-arxiv}.  \\

The gradient in surface tension is set up by the preferential evaporation of $\Omega_{\mathrm{v}}$ volume of IPA, at an evaporation rate $j_{\mathrm{v}}$, during this time $t^{\ast}$, given by
\begin{align}
	\Omega_{\mathrm{v}} \sim j_{\mathrm{v}} R^{\ast 2} t^{\ast} \sim \left( \phi_{0} \Omega_{0} - \phi_{\mathrm{c}} \Omega_{\mathrm{f}} \right) \, , \label{eq:ipa-evap}
\end{align}
where $\Omega_{0}$ and $\Omega_{\mathrm{f}}$ are the initial and final volumes of the mother droplet, respectively. Volume conservation of the non-volatile water component yields
\begin{align}
	\left( 1 - \phi_{0} \right) \Omega_{0} = \left( 1 - \phi_{\mathrm{c}} \right) \Omega_{\mathrm{f}} \, .
	\label{eq:water-conservation}
\end{align}
Combining equations \eqref{eq:ud-scale}, \eqref{eq:timescale}, \eqref{eq:ipa-evap}, and \eqref{eq:water-conservation}, we get the scaling relationships for $R^{\ast}$ and $t^{\ast}$, given by equations \eqref{eq:keiser-a} and \eqref{eq:keiser-b}, respectively. These scaling relationships provide a predictive framework to interpret the experimentally observed variations in droplet spreading and fragmentation timescales, particularly in the absence of viscoelastic effects. 

\newpage

\setcounter{equation}{0}
\setcounter{figure}{0}
\setcounter{table}{0}

\renewcommand{\thefigure}{S\arabic{figure}}
\renewcommand{\thetable}{S\arabic{table}}
\renewcommand{\theequation}{S\arabic{equation}}

\section{Supplementary information}

\FloatBarrier

\subsection{Density measurements}

Densities were determined by weighing at least three 100 \SI{}{\micro \text{L}} samples at 19 \SI{}{\text{\celsius}} using a precision analytical laboratory balance (Mettler-Toledo GmbH) with an accuracy of 0.1 mg. The density of sunflower oil ($\rho_{\mathrm{o}}$) was measured as 940 kg/m$^{\text{3}}$, while the density of the mother droplet ($\rho_{\mathrm{d}}$) remained effectively constant across polymer concentrations. Accordingly, for all analyses, $\rho_{\mathrm{d}}$ was approximated as 930 kg/m$^{\text{3}}$ for PEO1M and 940~kg/m$^{\text{3}}$ for PEO4M. 

\subsection{Interfacial tension measurements}

\begin{figure}
    \centering
    \includegraphics[width=\textwidth]{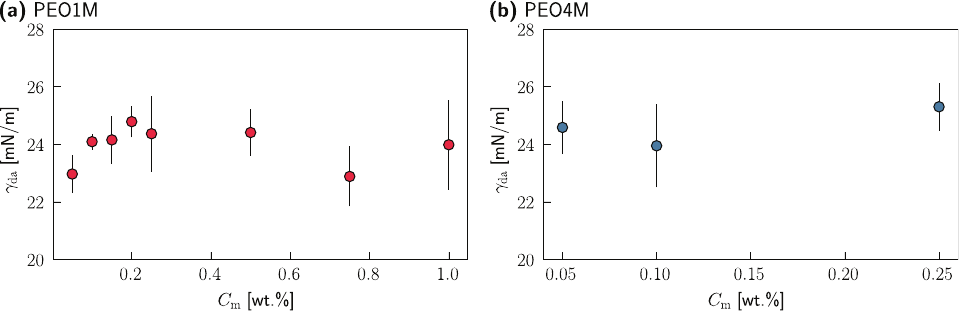}
     \caption{\textbf{Interfacial tension measurements.} Variation of drop-air interfacial tension coefficient, $\gamma_{\mathrm{da}}$, with polymer concentration, $C_{\mathrm{m}}$, for \textbf{(a)} PEO1M and \textbf{(b)} PEO4M. The discrete markers denote the mean of at least three independent experimental realizations while the error bars indicate $\pm$ one standard deviation.}
    \label{fig:fig-s1}
\end{figure}

Interfacial tension coefficients were obtained using the pendent drop method on a commercial drop shape analyzer (DSA 100E, Kr\"{u}ss GmbH). Images of axisymmetric droplets suspended from a hydrophobic Teflon-coated stainless steel needle (inner diameter = 0.25 mm, Nordson EFD) were recorded in both air and sunflower oil environments. The interfacial tension coefficients were extracted using the ``Pendent\_Drop" plugin \cite{daerr-2016-jopenressoftw} in the open-source image analysis software Fiji \cite{schindelin-2012-natmeth}. All measurements were performed at a temperature of 19 \SI{}{\text{\celsius}} and in triplicate. \\ 

The interfacial tension coefficient of the sunflower oil-air interface ($\gamma_{\mathrm{oa}}$) was determined to be 30.5 mN/m. Measurements (see figure \ref{fig:fig-s1}) indicated that both the drop-oil ($\gamma_{\mathrm{do}}$) and drop-air ($\gamma_{\mathrm{da}}$) interfacial tension coefficients were effectively independent of the polymer concentration, $C_{\mathrm{m}}$. Hence, we consider $\gamma_{\mathrm{da}}$~=~24~mN/m and $\gamma_{\mathrm{do}}$ = 4 mN/m for all calculations in the present study. 

\subsection{Rheological characterization}

Rheological measurements were performed on a stress-controlled rotational rheometer (MCR 501, Anton Paar GmbH) using a cone-and-plate geometry (1\SI{}{\text{\degree}} angle, 50 mm diameter, and mean gap of 0.1 mm). A solvent trap containing a 40\% (by mass) IPA-water solution was employed to prevent evaporation during the measurements. All measurements were performed at 19 \SI{}{\text{\celsius}} and in triplicate. \\ 

\begin{figure}
    \centering
    \includegraphics[width=\textwidth]{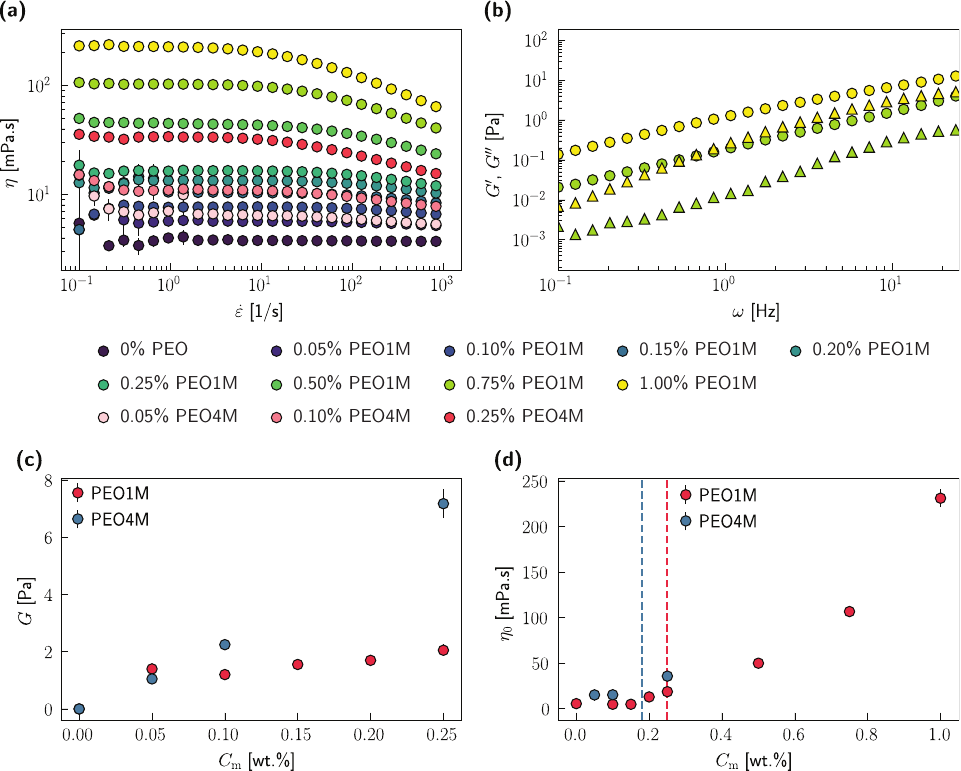}
     \caption{\textbf{Rheological characterization.} \textbf{(a)} Variation of shear viscosity, $\eta$, with shear rate, $\dot{\varepsilon}$, for different polymer concentrations. \textbf{(b)} Representative frequency ($\omega$) sweep measurements, at 10\% amplitude and for $C_{\mathrm{m}}$ = 0.75\% and 1.00\% PEO1M, where the triangular and circular data markers denote the storage ($G^{\prime}$) and loss ($G^{\prime \prime}$) moduli, respectively. Variations of \textbf{(c)} elastic modulus, $G$, and \textbf{(d)} zero-shear viscosity, $\eta_{0}$, with polymer concentration, $C_{\mathrm{m}}$. The dashed lines in panel d denote the estimates of the entanglement concentration, $C_{\mathrm{e}}$, for both PEO1M (red) and PEO4M (blue). In panels a, c, and d, the discrete markers denote the mean of at least three independent experimental realizations while the error bars indicate $\pm$ one standard deviation.}
    \label{fig:fig-s2}
\end{figure}

The shear viscosity ($\eta$) as a function of the shear rate ($\dot{\varepsilon}$) for different polymer concentrations ($C_{\mathrm{m}}$) is presented in figure \ref{fig:fig-s2}a. The focus of the present work is on elucidating the role of viscoelasticity on the Marangoni bursting behavior. However, dissolved polymers may also impart shear thinning behavior (as also observed at high $C_{\mathrm{m}}$ in figure \ref{fig:fig-s2}a), which further complicates the dynamics of an already complex phenomenon. Hence, in the present study, we limit $C_{\mathrm{m}}$ such that the polymeric liquids behave predominantly as Boger fluids \cite{james-2009-arfm}, i.e. their shear viscosity is independent of the shear rate (although there is departure from Boger fluid-like behavior at high $C_{\mathrm{m}}$ for high shear rates, as seen in figure \ref{fig:fig-s2}a). The corresponding zero-shear viscosities ($\eta_{0}$) can also be extracted from the shear viscosity vs. shear rate curves for the different polymer concentrations. \\

Frequency ($\omega$) sweep measurements were performed at an amplitude of 10\% (determined after thorough amplitude sweep measurements) to determine the linear viscoelastic storage ($G^{\prime}$) and loss ($G^{\prime \prime}$) moduli of the polymeric liquids used in the present work. All measurements were performed at 19 \SI{}{\text{\celsius}}. The measurements for two representative concentrations ($C_{\mathrm{m}}$ = 0.75\% and 1.00\% PEO1M) are shown in figure \ref{fig:fig-s2}b. While no crossover between $G^{\prime}$ and $G^{\prime \prime}$ is observed within the frequency ($\omega$) range shown in figure \ref{fig:fig-s2}b, the experimental trends suggest the crossover frequency to be $\mathcal{O}(\text{100} \, \text{Hz})$---consistent with prior measurements \cite{sen-2022-arxiv, li-2025-arxiv} of relaxation times of PEO solutions of comparable concentrations. However, at such high frequencies, inertial artifacts in the frequency sweep measurements became unavoidable, resulting in unreliable data. Hence, extensional thinning measurements were relied upon to accurately determine the relaxation times and elastic moduli of the polymeric solutions. \\

The relaxation times of the polymer solutions ($\tau$) were measured from the extensional thinning of liquid filaments in a pendent droplet configuration \cite{deblais-2018-prl, sur-2018-jor, mathues-2018-jor, deblais-2020-jfm}, which is known \cite{sen-2021-jfm, sen-2022-arxiv, li-2025-arxiv} to provide accurate estimates of the relaxation times for stretching filaments of PEO solutions. All measurements were performed at 19 \SI{}{\text{\celsius}} and in triplicate. Knowing the zero-shear viscosity ($\eta_{0}$, which is also the shear viscosity at all shear rates for Boger fluids) and the relaxation time ($\tau$) for different polymer concentrations allows for the estimation of the elastic modulus: $G = \eta_{0} / \tau$. The variation of the elastic modulus ($G$) with the polymer concentration ($C_{\mathrm{m}}$) is shown in figure \ref{fig:fig-s2}c. The elastic modulus ($G$) increases with increasing polymer concentration ($C_{\mathrm{m}}$) for both PEO1M and PEO4M, with steeper increases observed for the higher molecular weight polymer (PEO4M).  \\

The entanglement concentration ($C_{\mathrm{e}}$) is determined to be the polymer concentration at which the zero-shear viscosity ($\eta_{0}$) of the polymer solution rapidly increases \cite{heo-2005-jor}. The variation of zero-shear viscosity ($\eta_{0}$) with polymer concentration ($C_{\mathrm{m}}$) is shown in figure \ref{fig:fig-s2}d. For the polymers used in the present study, the entanglement concentrations were determined to be $\approx$ 0.25\% (by mass) and 0.18\% (by mass) for PEO1M and PEO4M, respectively (dashed lines in figure \ref{fig:fig-s2}d). 

\subsection{Spreading dynamics, droplet lifetime, and fragmentation time}

\begin{figure}
    \centering
    \includegraphics[width=\textwidth]{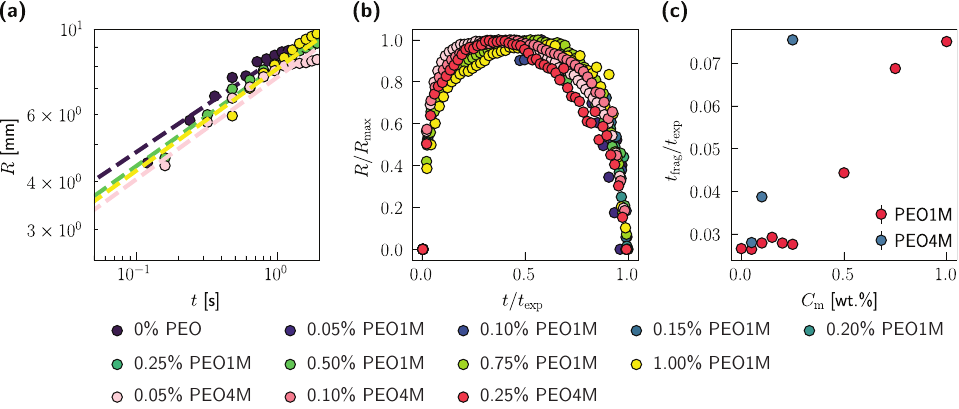}
     \caption{\textbf{Spreading dynamics, normalized spreading, and fragmentation times.} \textbf{(a)} Early time spreading dynamics of the mother droplet, where the dashed lines denote $R \sim t^{1/4}$. \textbf{(b)} Variation of normalized spreading radius, $R/R_{\max}$, with normalized spreading time, $t/t_{\exp}$, for different polymer concentrations, where all experimental datapoints tend to collapse on one master curve. \textbf{(c)} Variation of the normalized fragmentation time, $t_{\mathrm{frag}}/t_{\exp}$, with polymer concentration, $C_{\mathrm{m}}$, which exemplifies the relative delay in self-emulsification as the polymer concentration increases. The discrete markers in panel c denote the mean of at least three independent experimental realizations while the error bars indicate $\pm$ one standard deviation.}
    \label{fig:fig-s3}
\end{figure}

The dynamics of Marangoni-stress driven spreading processes can be captured from simple scaling arguments \cite{jensen-1995-jfm, afsarsiddiqui-2003-langmuir, chan-2024-pnasnexus}. In the present work, a water-IPA droplet (with or without dissolved polymers, since the presence of dissolved polymers were observed to not have any significant impact on the spreading dynamics; see figure~\ref{fig:fig-1}a), with IPA mass fraction $\phi$, spreads on an oil layer of density $\rho_{\mathrm{o}}$, with an instantaneous spreading radius of $R(t)$. The alcohol surface concentration, $\Gamma$, scales as \cite{jensen-1995-jfm, chan-2024-pnasnexus} $\Gamma \sim \phi / R^{2}$, resulting in a Marangoni stress given by \cite{chan-2024-pnasnexus}
\begin{align}
    \sigma_{\mathrm{Marangoni}} \sim \frac{A \Gamma}{R} \sim \frac{A \phi}{R^{3}} \, ,
    \label{eq:si-marangoni-stress}
\end{align}
where $A = - \mathrm{d} \gamma / \mathrm{d} \Gamma$ represents the gradient of surface tension, $\gamma$, with surface concentration, and can be considered to be equivalent to a surface activity \cite{chan-2024-pnasnexus}. This spreading liquid front experiences a viscous resistance from the oil substrate, and the corresponding viscous stress is given by 
\begin{align}
    \sigma_{\mathrm{viscous}} \sim \frac{\eta_{\mathrm{o}} R}{t \, h_{\mathrm{o}}} \, ,
    \label{eq:si-viscous-stress}
\end{align}
where the viscous boundary layer is considered to penetrate the entire oil layer thickness, $h_{\mathrm{o}}$, at time scales much faster than the characteristic spreading time scale (see also the derivation for equation \eqref{eq:ud-scale}). If the alcohol concentration, $\phi$, is sufficiently large, as in the present case, $A$ behaves as a constant in time. Hence, combining equations \eqref{eq:si-marangoni-stress} and \eqref{eq:si-viscous-stress} results in
\begin{align}
    R \sim t^{1/4} \, ,
    \label{eq:si-r-scaling}
\end{align}
which has also been observed in similar experimental configurations \cite{hack-2024-jcis, chan-2024-pnasnexus}. The trends described by the dashed lines in figure \ref{fig:fig-s3}a are also consistent with this $R \sim t^{1/4}$ scaling, but given the limited range of our data, we refrain from claiming any scaling relationship at this point. \\

The variation of the normalized spreading radius, $R/R_{\max}$, with normalized spreading time, $t/t_{\exp}$, is shown in figure \ref{fig:fig-s3}b for different polymer concentrations. All experimental datapoints tend to collapse on one master curve, which was also observed in prior studies with Newtonian fluids \cite{keiser-2017-prl}. \\

The variation of the normalized fragmentation time, $t_{\mathrm{frag}}/t_{\exp}$, with polymer concentration, $C_{\mathrm{m}}$, is shown in figure \ref{fig:fig-s3}c. The normalized fragmentation time increases with increasing $C_{\mathrm{m}}$, thus emphasizing the relative delay in self-emulsification, and thus the extended lifetime of the mother droplet, as the polymer concentration increases. 

\subsection{Number of fingers decreases with increasing polymer concentration}

\begin{figure}
    \centering
    \includegraphics[width=\textwidth]{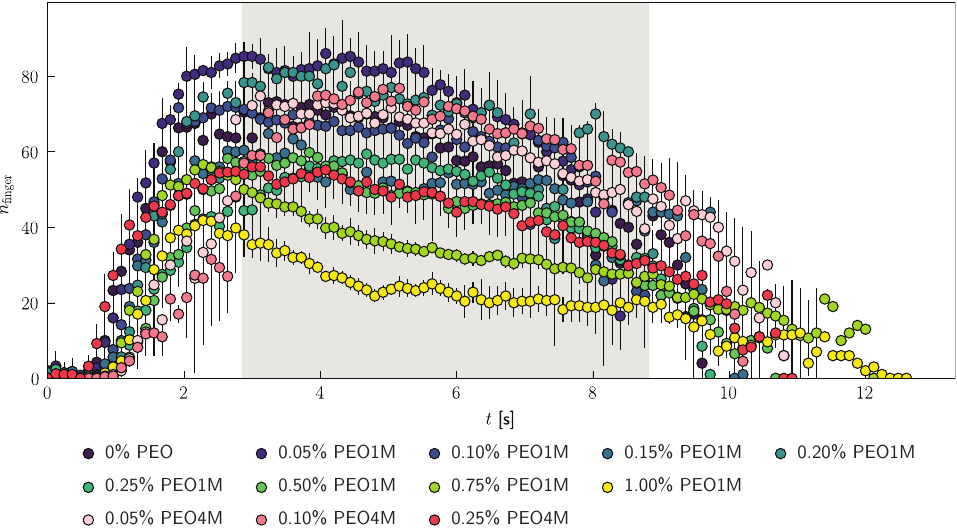}
    \caption{\textbf{Number of fingers.} Temporal variation of the number of fingers, $n_{\mathrm{finger}}$, for different polymer concentrations. The shaded area denotes the time-span corresponding to stable finger detection. The discrete markers denote the mean of at least three independent experimental realizations while the error bars indicate $\pm$ one standard deviation.}
    \label{fig:fig-s4}
\end{figure}

The temporal variation of the mean finger number, $n_{\mathrm{finger}}$, for different polymer concentrations is shown in figure \ref{fig:fig-s4}. To filter out the misdetection of fingers, we only consider experimental snapshots containing at least 10 fingers. Moreover, for $t \lessapprox$ 3 s, the finger dimensions fall below the reliable detection limit of the image processing algorithm, thus underestimating the number of fingers. Consequently, we observe a false convergence of $n_{\mathrm{finger}}$ towards 0 at time $t = 0$. However, our experiments indicate that the fingering instability develops almost instantaneously. Hence, figure \ref{fig:fig-s4} underestimates the number of fingers for $t \lessapprox$ 3 s. A similar underestimation was also observed towards the end of each experiment, i.e. for $t \gtrapprox$ 8 s. To mitigate this, we focus on the time-span where the finger detection is robust, denoted by the shaded area in figure \ref{fig:fig-s4}. Within this time-span, the number of fingers gradually decreases with time for all polymer concentrations. However. since the spreading radius, $R$, of the mother droplet also decreases within this time-span (as seen in figure \ref{fig:fig-2}a), the instability wavelength, $\lambda = 2 \pi R / n_{\mathrm{finger}}$, remains fairly constant (as seen in figure \ref{fig:fig-3}a). Additionally, the number of fingers is observed to decrease with increasing polymer concentration. 

\subsection{Viscoelastic control of self-emulsification is polymer-independent}

The broader applicability of this method of leveraging polymer-induced viscoelasticity to control Marangoni-driven self-emulsification requires the method to be independent of the molecular structure of the dissolved polymer (i.e. polyethylene oxide (PEO) in the main manuscript). This was confirmed by performing Marangoni-bursting experiments with polyvinylpyrrolidone (average molecular weight $\approx$ 360 $\times$ 10$^{3}$ Da, Sigma-Aldrich, henceforth referred to as ``PVP") as the dissolved polymer, instead of PEO. The IPA concentration in the resulting polymeric solution (of IPA, water, and PVP) was kept constant at 40\% (by mass) while the PVP concentration ($C_{\mathrm{PVP}}$) was varied between 1.0\% and 5.0\%. The materials characterization methods, experimental protocols, and analysis techniques were identical to the ones described for PEO solutions in the main manuscript (see movie SM5 for typical experimental realizations of Marangoni-driven self-emulsification with PVP solutions).  \\ 

\begin{figure}
    \centering
    \includegraphics[width=0.89\textwidth]{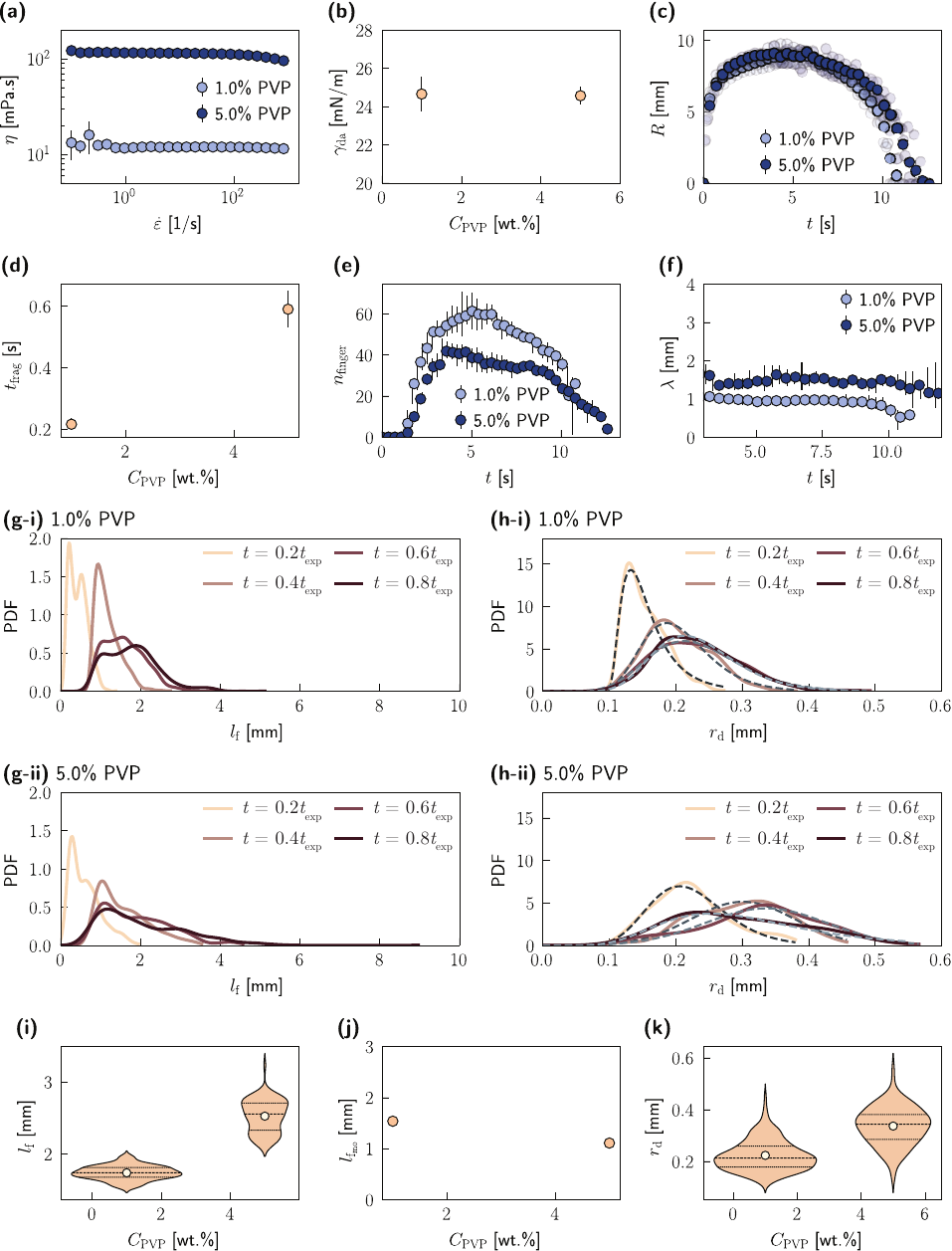}
    \caption{\textbf{Marangoni-driven self-emulsification of PVP-containing droplets.} \textbf{(a)} Variation of shear viscosity, $\eta$, with shear rate, $\dot{\varepsilon}$, for two different PVP concentrations. \textbf{(b)} Variation of the drop-air interfacial tension coefficient, $\gamma_{\mathrm{da}}$, with PVP concentration, $C_{\mathrm{PVP}}$. \textbf{(c)} Temporal evolution of the spreading front radius, $R$, for two different PVP concentration. The transulucent symbols denote at least three independent experimental realizations per polymer concentration while the opaque symbols indicate the mean for each polymer concentration. \textbf{(d)} Time of incipence of self-fragmentation, $t_{\mathrm{frag}}$, for two different PVP concentrations. Temporal variation of \textbf{(e)} the number of fingers, $n_{\mathrm{finger}}$, and \textbf{(f)} the instability wavelength, $\lambda$, for two different PVP concentrations. The discrete markers in panels a, b, and d -- f denote the mean of at least three independent experimental realizations while the error bars indicate $\pm$ one standard deviation. Distributions of the \textbf{(g)} finger length, $l_{\mathrm{f}}$, and \textbf{(h)} daughter droplet radii at different time instants, and for two different PVP concentrations. The dashed lines in panels h-i and h-ii denote the corresponding log-normal distribution fits. \textbf{(i)} Distributions of finger length, $l_{\mathrm{f}}$, at $t = 0.5 \, t_{\exp}$ for two different PVP concentrations. \textbf{(j)} Variation of the characteristic finger length, $l_{\mathrm{f_{\mathrm{mo}}}}$, at $t = 0.6 \, t_{\exp}$ for two different PVP concentrations. \textbf{(k)} Distributions of daughter droplet radius, $r_{\mathrm{d}}$, at $t = 0.5 \, t_{\exp}$ for two different PVP concentrations. The discrete datapoints in panels i and k denote the mean value of each distribution. See movie SM5 for the corresponding movies. }
    \label{fig:fig-s5}
\end{figure}

The densities of the PVP solutions were measured to be 936 and 953 kg/m$^{\text{3}}$ for the 1.0\% and 5.0\% concentrations, respectively. Extensional thinning measurements indicated that while the 5.0\% PVP solution had a relaxation time ($\tau$) of 1.23 ms, no prominent elastocapillary thinning was observed for the 1.0\% PVP solution. Hence, the 1.0\% PVP solution was adjudged to have a Newtonian response. Shear rheology (see figure \ref{fig:fig-s5}a) also indicated that both the 1.0\% and 5.0\% PVP solutions behaved as Boger-like fluids, although a slight shear-thinning was observed at high shear rates for the 5.0\% PVP solution. Measurements (see figure \ref{fig:fig-s5}b) further confirmed the drop-oil ($\gamma_{\mathrm{da}}$) and drop-air ($\gamma_{\mathrm{da}}$) interfacial tension coefficients to be effectively independent of the polymer concentration, $C_{\mathrm{PVP}}$, at $\gamma_{\mathrm{da}}$ = 24.5 mN/m and $\gamma_{\mathrm{do}}$ = 4 mN/m, respectively. \\

The temporal evolution of the spreading front radius, $R$, follows similar dynamics as PEO-containing droplets (see figures \ref{fig:fig-s5}c and \ref{fig:fig-2}a)---first increasing till reaching a maximum and subsequently decreasing, the latter being accompanied by a spontaneous destabilization of the droplet periphery, resulting in self-emulsification (see movie SM5 for the corresponding movies). Here as well, the introduction of viscoelasticity---achieved via the increment of the concentration of dissolved PVP---delays the onset of self-emulsification, as shown in figures \ref{fig:fig-s5}d, \ref{fig:fig-2}d, and \ref{fig:fig-s3}c.  \\ 

Similar to the PEO-containing droplets, fragmentation here also proceeds with the formation of protruding fingers from the periphery of the spreading droplet, where the number of fingers ($n_{\mathrm{finger}}$) first increases with time, and then gradually decreases (see figures \ref{fig:fig-s5}e and \ref{fig:fig-s4}). However, the finger detection algorithm also suffers from the same shortcoming at early times ($t \lessapprox$ 3 s) as the one described for PEO-containing droplets (figure \ref{fig:fig-s4}), resulting in an underestimation of the number of fingers at early times. The number of fingers was also observed to decrease with increasing PVP concentration, mirroring the trend observed for PEO-containing droplets (see figures \ref{fig:fig-s5}e and \ref{fig:fig-s4}). A similar correspondence between the PEO- and PVP-containing droplets was also observed for the temporal variation of the instability wavelength, $\lambda$, which remains practically constant in time, but this constant value increases with increasing polymer concentration (see figures \ref{fig:fig-s5}f and \ref{fig:fig-3}a).  \\

The finger length, $l_{\mathrm{f}}$, distribution exhibit a gradual broadening with time, as seen in figures \ref{fig:fig-s5}g-i and \ref{fig:fig-s5}g-ii. The broadening effect is also more pronounced at higher PVP concentrations. Moreover, both of these observations are consistent with our observations for PEO-containing droplets (see figures \ref{fig:fig-4}a-i -- \ref{fig:fig-4}a-iii). The similarities with PEO-containing droplets in the temporal evolution of distributions also extend to the daughter droplet radius ($r_{\mathrm{d}}$) distribution (see figures \ref{fig:fig-s5}h-i and \ref{fig:fig-s5}h-ii), where a shift towards larger droplet sizes at later times is also observed. The daughter droplet radius distributions are also well-fitted by the log-normal distribution here (dashed lines in figures \ref{fig:fig-s5}h-i and \ref{fig:fig-s5}h-ii), consistent with the PEO-containing droplets (see figures \ref{fig:fig-5}a-i -- \ref{fig:fig-5}a-iii). The variations of the distributions of $l_{\mathrm{f}}$ (figure \ref{fig:fig-s5}i) and $r_{\mathrm{d}}$ (figure \ref{fig:fig-s5}k) with PVP concentration ($C_{\mathrm{PVP}}$) also show the same trend as PEO-containing droplets (see figures \ref{fig:fig-4}b-i, \ref{fig:fig-4}b-ii, \ref{fig:fig-5}b-i, and \ref{fig:fig-5}b-ii). Hence, we can conclude that leveraging polymer-induced viscoelasticity to control the dynamics of Marangoni-driven self-emulsification at fluid-fluid interfaces is independent of the molecular structure of the dissolved polymer.

\subsection{Supplementary movies}

\textbf{Movie SM1:} Marangoni-driven self-emulsification of a water-IPA droplet containing 0.10\% PEO1M. \\

\textbf{Movie SM2:} Marangoni-driven self-emulsification of a water-IPA droplet containing 0.50\% PEO1M. \\

\textbf{Movie SM3:}  Spreading dynamics during Marangoni-driven self-emulsification of a water-IPA droplet ($C_{\mathrm{m}}$~=~0\%). \\

\textbf{Movie SM4:} Marangoni-driven self-emulsification of a water-IPA droplet containing 0.90\% PEO4M. \\

\textbf{Movie SM5:} Marangoni-driven self-emulsification of water-IPA droplets containing 1.0\% and 5.0\% PVP.

\newpage

\bibliographystyle{unsrt}
\bibliography{marangoni-bursting.bib}


\end{document}